\documentclass[final,1p,times]{elsarticle}

\usepackage{amsmath,amssymb,graphicx,color,bm,epstopdf,float,caption,subcaption}

\journal{Superlattices and Microstructures}

\begin{document}

\title{Tuning of Zero Energy States in Quantum Dots of Silicene and  Bilayer Graphene  by Electric Field}

\author[1,2]{Hazem Abdelsalam \corref{cor1}}

\address[1]{University of Picardie, Laboratory of Condensed Matter
Physics, Amiens, 80039, France} 
\address[2]{Department of Theoretical Physics, National Research Center, Cairo,12622, Egypt}
\cortext[cor1]{corresponding author}
\ead{hazem.abdelsalam@etud.u-picardie.fr}

\author[3]{T. Espinosa-Ortega}
\address[3]{Division of Physics and Applied Physics, Nanyang
Technological University 637371, Singapore}
\author[1,4]{Igor Lukyanchuk}
\address[4]{ L. D. Landau Institute for Theoretical Physics, Moscow, Russia}

\begin{abstract}

Electronic properties of triangular and hexagonal nano-scale quantum dots (QDs) of Silicene and bilayer graphene are studied. It is shown that the low-energy edge-localized electronic states, existing within the size-quantized gap are easily tunable by electric field. The appearance and field evolution of the electronic gap in these zero energy states (ZES) is shown to be very sensitive to QD geometry that permits to design the field-effect scalable QD devices with  electronic properties on-demand.

\end{abstract}


\maketitle



\section{Introduction}

\label{sec:intro}

\begin{figure}[htp]
\centering \includegraphics[width=0.7\textwidth]{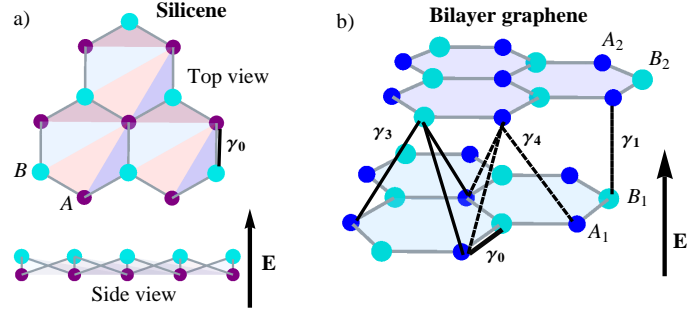}
\caption{ Structure and coupling parameters for  Silicene QD (a) and for graphene bilayer QD (b).}
\label{Fig1}
\end{figure}

Interest to new graphene-like materials is related with the rising quest to
develop the nano-scale field-effect transistor \cite{Ni,Tao}, unifying the remarkable
electronic properties of graphene with possibility of easy tuning by
electric field. Given that the monolayer graphene itself is not quite
sensitive to the applied field, the natural way consists in creations of the
multi-layer structure with gate-controlled potential difference between
layers. Two systems are promising: the artificial monolayer materials, like
Silicene, Germanene, etc. \cite%
{Tsai2013,Bianco2013,CLiu2011_2,Bechstedt} and
bilayer \cite{McCann} (and in general multilayer) graphene structures. The principal distinction of the first group
(we consider Silicene for definitiveness) 
is their buckled structure that separates A and B atoms of the honeycomb lattice in
the transversal direction (Fig. \ref{Fig1}a) and provides the required 
gradient of potential. As a result, the band structure can be controlled by electric
field that tunes the gap and induces  transition from a topological
insulator to a band insulator \cite{CLiu2011,Drummond,Ezawa,Tsaran}. The field-provided interlayer potential
difference in bilayer graphene (Fig. \ref{Fig1}b) also opens a gap between the
conduction and the valence bands \cite{McCann,Castro2007}, controllable by transistor gate.

Reduction of the lateral dimension of the discussed systems to the size of nanoscale
quantum dot (QD) changes however their electronic properties leading to
breakdown of the band structure and to enhancement of the role of the edge
electronic states \cite{Ezawa3,Zhang2008,Rossier2007,Wang2009,Potasz,Portnoi,Zarenia,Espinosa,Espinosa2}. Recent studies demonstrated that the electronic
properties of  QDs in graphene-type materials
\cite{,Guclu,Hazem}
are provided by distribution of the localized edge states in the low-energy
spectral region. Depending on the geometrical parameters such as size,
shape, edge termination and number of layers  the situation can
be drastically changed from the uniformly-distributed edge-localized states  to the low-energy size-quantized gap with central highly-degenerate
peak of zero energy states (ZES) in the middle \cite{Hazem}. 

In this letter we investigate how the electronic states in Silicene and
bilayer graphene QDs can be tuned by the transversal electric field and what
impact on the future nanoscale field-effect device engineering can be expected. For
calculations we use the standard tight-binding model for clusters with about
450 atoms per layer. We select the most indicative geometries of triangular
and hexagonal QDs with zigzag edge termination.


\section{Model and Density of States}

\label{sec:Model}

The electronic properties of graphene-type materials in transversal electric
field can be calculated using the tight-binding Hamiltonian \cite{  McCann,CLiu2011_2},

\begin{equation}
H=\sum_{\left\langle ij\right\rangle }t_{ij}c_{i}^{\dag
}c_{j}+\sum_{i}V_{i}\left( E\right) c_{i}^{\dag }c_{i}  \label{Htb_s}
\end{equation}%
where $c_{i}^{\dag }$ and $c_{i}$ are the electron creation and annihilation
operators, $t_{ij}$ are the inter-site hopping parameters and $V_{i}$ is the
on-site electron potential that depends both on the local atomic environment
and on the applied electric field. In cases of Silicene and bilayer graphene the parameters $t_{ij}$ can be written via the
nearest neighbor (NN) coupling constants $\gamma _{i}$, as shown in Fig. 1.

Specifying Hamiltonian (\ref{Htb_s}) for the case of Silicene we use the simplified version, appropriate for the low-energy states \cite{Ezawa,CLiu2011_2}.  In this approximation there is only one \textit{in-plane} coupling parameter between sites A and B, $\gamma
_{0}\simeq 1.6~\mathrm{eV}$, whereas
the on-site potential,  $V_{i}(E)$ is different for A and B sites and can be presented as  $V_{i}=\xi _{i}\Delta -\xi _{i}lE$ where $\xi _{i}=\pm 1$ for
the B and A type of atoms, $\ \Delta \simeq 3.9~\mathrm{meV}$ is the
effective buckling-gap parameter and $lE$ is the field-provided
electrostatic interaction, related to the up/down shift of B and A atoms on $%
l\simeq 0.23\,\mathring{A}$ with respect to the average plane.

For graphene bilayer structure, besides the \textit{in-plane} coupling $%
\gamma _{0}\simeq 3.16~\mathrm{eV}$ the \textit{interlayer} parameters $%
\gamma _{1}\simeq 0.38~\mathrm{eV}$, $\gamma _{3}\simeq 0.38~\mathrm{eV}$
and $\gamma _{4}\simeq 0.14~\mathrm{eV}$ (Fig. \ref{Fig1}b) should be also taken into
account. The field-dependent on-site potential can be written as $%
V_{i}=\eta _{i}\Delta -\varsigma _{i}lE$ \cite{ McCann} were $\eta _{i}=0$ for A1 and B2
atoms, $\eta _{i}=1$ for A2 and B1 atoms and $\varsigma _{i}=\pm 1$ for
atoms, located in the upper (A2, B2) and lower (A1, B1) layers
correspondingly (See Fig.$~$1). The site-environment gap parameter is taken
as $\Delta \simeq 22~\mathrm{meV}$  and the interlayer distance as $%
2l \simeq 3.5\,\mathring{A}$.

\begin{figure}[htp]
\centering
\includegraphics[height=5.5cm,width=.7\textwidth]{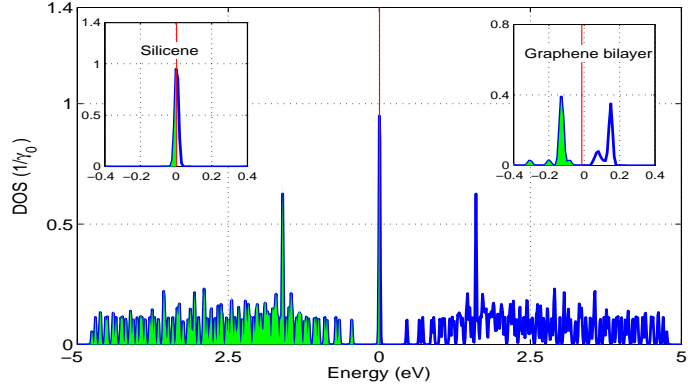}
\caption{DOS for triangular QDs for graphene-like materials at zero field.
Insets show the low-energy zoom of ZES central peaks for Silicene (left) and
for Graphene bilayer (right). Here and further the vertical red line
demarcates the Fermi level. The filled electronic states are shaded by green
color.}
\label{Fig2}
\end{figure}

The density of states (DOS) for triangular clusters of
Silicene and bilayer graphene obtained at $E=0$ by numerical diagonalization
of Hamiltonian (\ref{Htb_s}) is shown in Fig. \ref{Fig2} where the discrete
electronic levels were convoluted with Gaussian $e^{-\varepsilon ^{2}/\delta
^{2}}$ (with $\delta \simeq 14~\mathrm{meV}\simeq 160~\mathrm{K}$) that
models the temperature and inhomogeneity-provided smearing. On large energy
scale, the obtained DOS is similar to that for the graphene clusters \cite{Espinosa}. The
two-peak band envelope of infinite-graphene DOS is spotted by the
finite-quantization cusps that vanishes with increasing of the cluster
size.\ Most importantly, the gap in electronic states is observed at
near-zero energies with sharp, almost degenerate central peak, located
inside the gap. This feature is provided by the edge-localized zero energy
states (ZES), that are bunched in the middle of the gap. The number of ZES, $%
\eta _{0}$ is as large as disbalance between A
and B . It reaches the maximum in case of triangular QD, for which the A-B
disbalance exactly corresponds to the number of ZES and is related to the
total number of atoms in QD $N$ as $\eta _{0}^{\Delta }=\sqrt{N+3}-3$ \cite{Rossier2007,Potasz,Espinosa2}.
Then, the gap is inversely proportional to $\eta _{0}^{\Delta }$ and can be
expressed via the in-plane coupling constant $\gamma _{0}$ as $2\Delta \simeq
11.12\gamma _{0}/\eta _{0}^{\Delta }$ \cite{Espinosa2}.

The fine structures of DOS for ZES in triangular QDs of Silicene and bilayer
graphene are shown on insets to Fig. \ref{Fig2}. In Silicene, like in single-layer
graphene, all ZES are concentrated exactly at zero energy, being accumulated
into the highly degenerate state with degeneracy factor $\eta _{0}^{\Delta }$%
. In contrast, the central peak of ZES in bilayer graphene is smeared by the
NN\ interlayer coupling parameter $\gamma _{4}$ with formation of the
finite-width double-peak structure. It is worth to note that oftenly only the
non-smearing  coupling $\gamma _{1}$ is taken into account and constrain $%
\gamma _{3,4}=0$ is assumed. Then the splitting due to $\gamma _{4}$ is
overlooked and ZES is still considered to be highly degenerated. \ 

For another QD geometries the A-B disbalance vanishes. Thus, the number of A
and B atoms in hexagonal QD is equal, $\eta _{0}^{hex}=0$ and  the
edge-localized states are uniformly dispersed within the size-quantized gap.

In next Chapter we shall study how these prominent features  evolve with application of the electric
field.

\section{Field Tuning of Zero Energy States}

\subsection{Silicene}

\begin{figure}[tbp]
\centering
\begin{subfigure}{.6\textwidth}
  \centering
  \includegraphics[width=\textwidth]{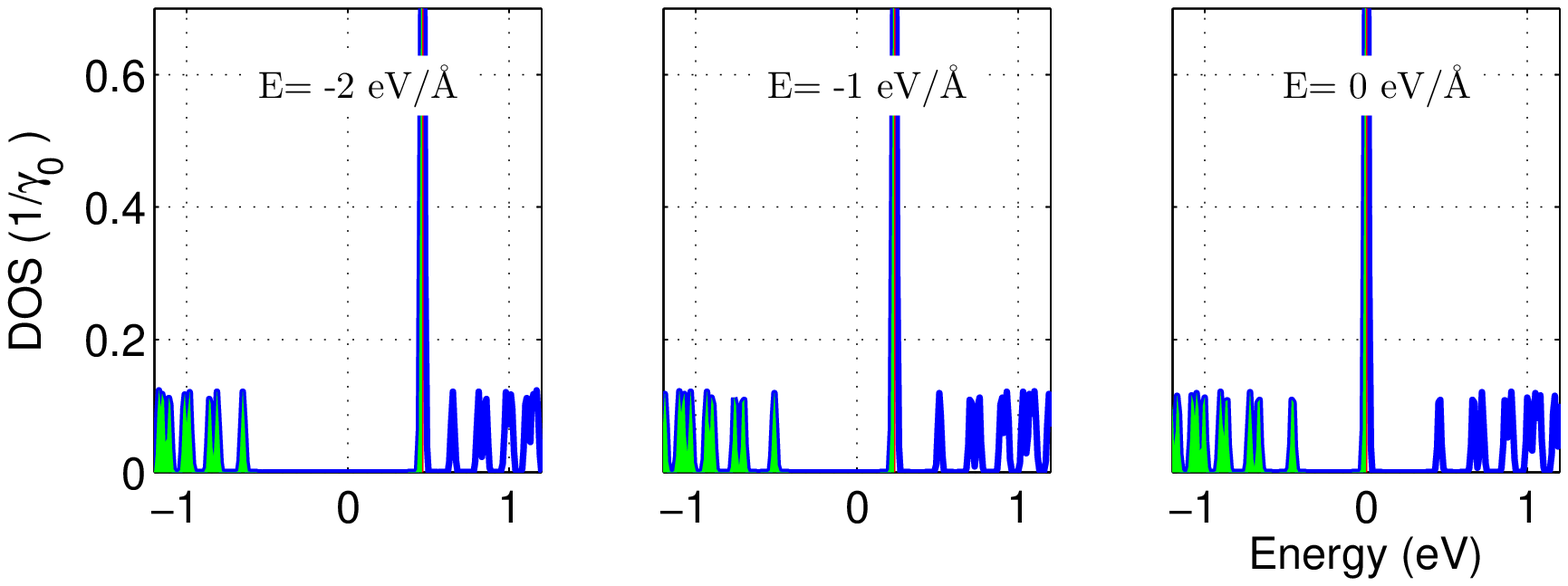}
  
\end{subfigure}%
\begin{subfigure}{.4\textwidth}
  \centering
 \hspace{.4cm}  \vspace*{.8cm}\centerline{\includegraphics[width=.9\textwidth]{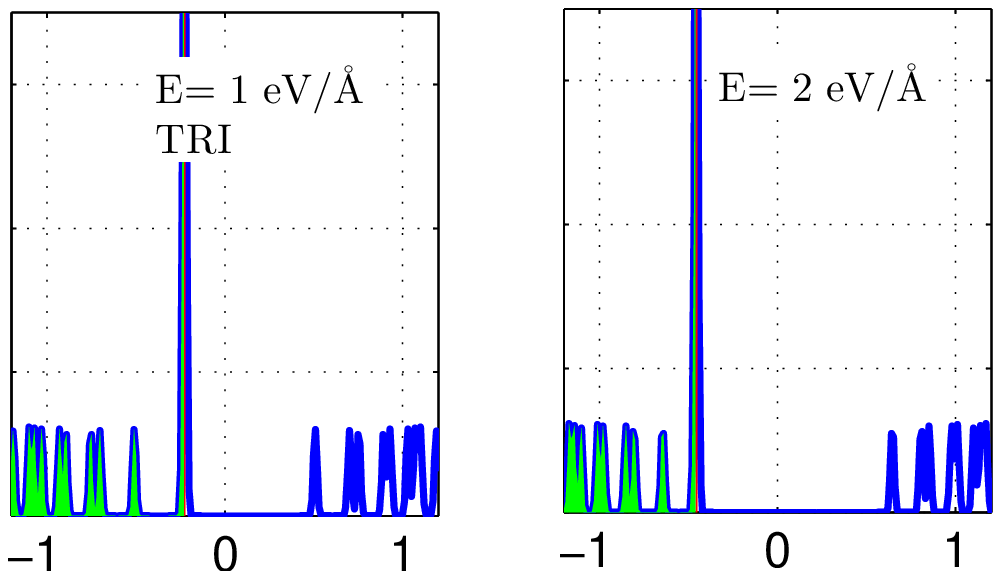}}
\end{subfigure}
\\[-0.5cm]
\begin{subfigure}{.6\textwidth}
  \centering
  \hspace{-.5cm}\includegraphics[width=\textwidth]{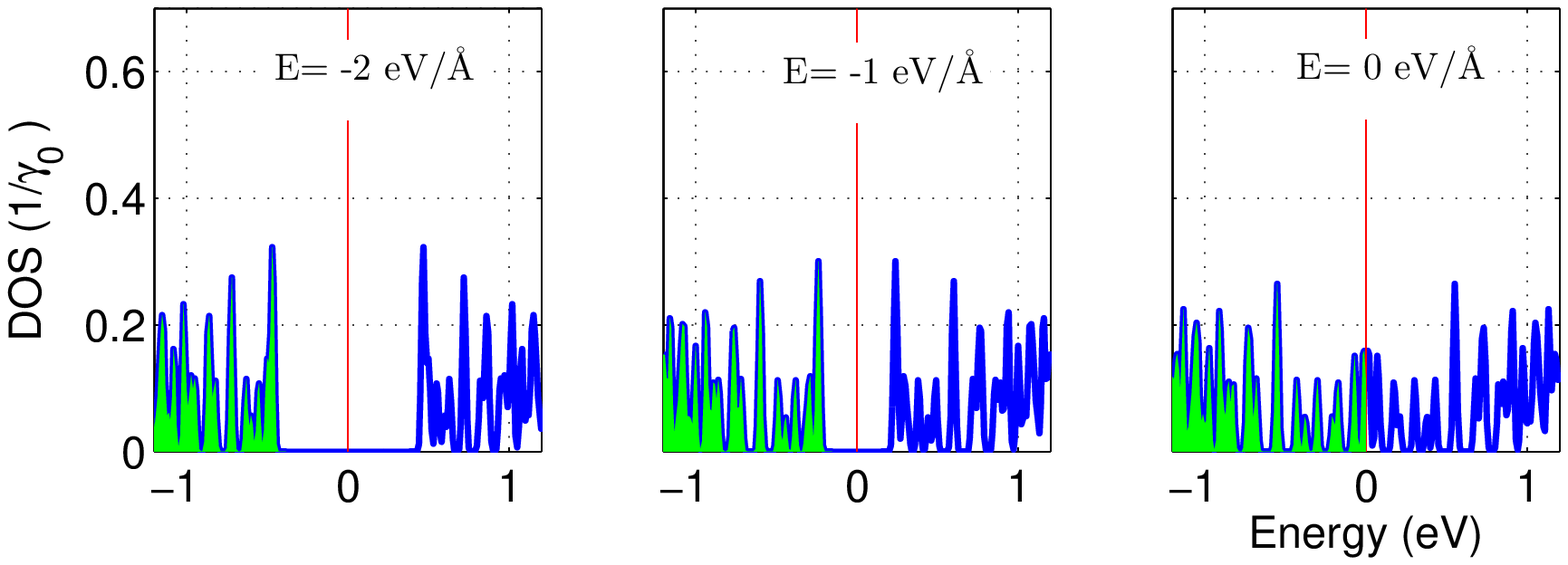}
\end{subfigure}%
\begin{subfigure}{.4\textwidth}
  \centering
 \hspace{2.5cm}  \vspace*{.25cm} \centerline{\includegraphics[height=2.8cm,width=.9\textwidth]{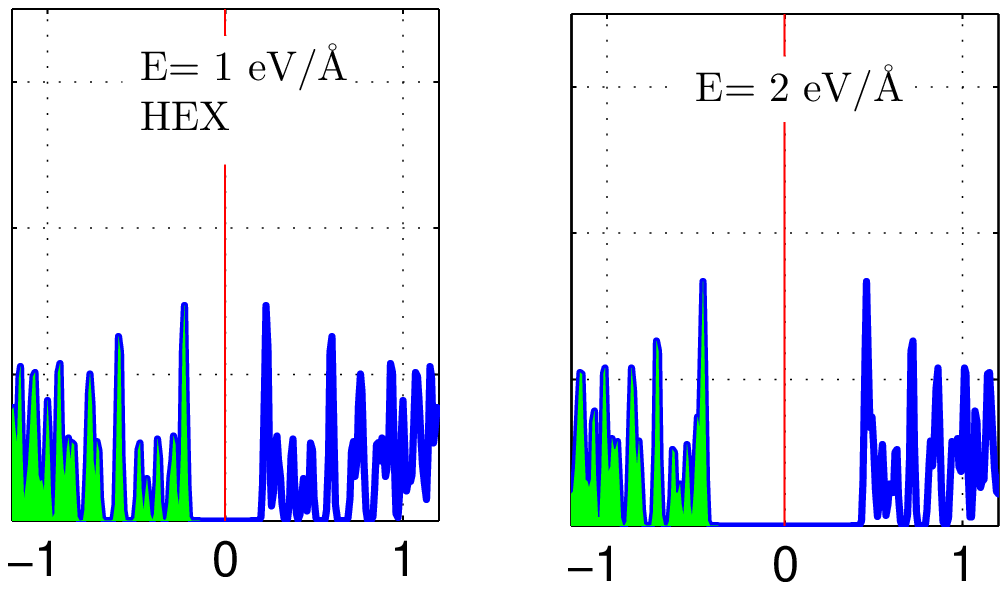}}
\end{subfigure}
\caption{Evolution of ZES for triangular (upper panel) and for hexagonal
(low panel) Silicene QDs as function of the applied electric field.}
\label{Fig3}
\end{figure}

Different number of up-shifted A atoms and down-shifted B atoms in
triangular Silicene QD breaks the mirror cluster symmetry and results in the different reaction of QD on the up- and
down-oriented electric field. The corresponding evolution of ZES under the
field application (Fig. \ref{Fig3}, upper panel) demonstrates that electric field
uniformly displaces the degenerate ZES level through the gap from 
conducting to  valence band by crossing zero energy level  at $E=0$. Having
the Fermi level pinned by ZES peak this gives the excellent possibility to
manipulate the electron interband hopping by electric field.

In contrast, the number of A and B atoms in hexagonal QD is equal and
initially the size-quantized gap spectrum range is filled by dispersed ZES.
Application of electric field opens the gap and symmetrically extends it to
valence and conducting bands (Fig. \ref{Fig3}, lower panel) enabling again the
efficient tuning of the electronic and optical properties of the system.

\subsection{Bilayer graphene}

\begin{figure}
\centering
\begin{subfigure}{.6\textwidth}
  \centering
    \includegraphics[width=\textwidth]{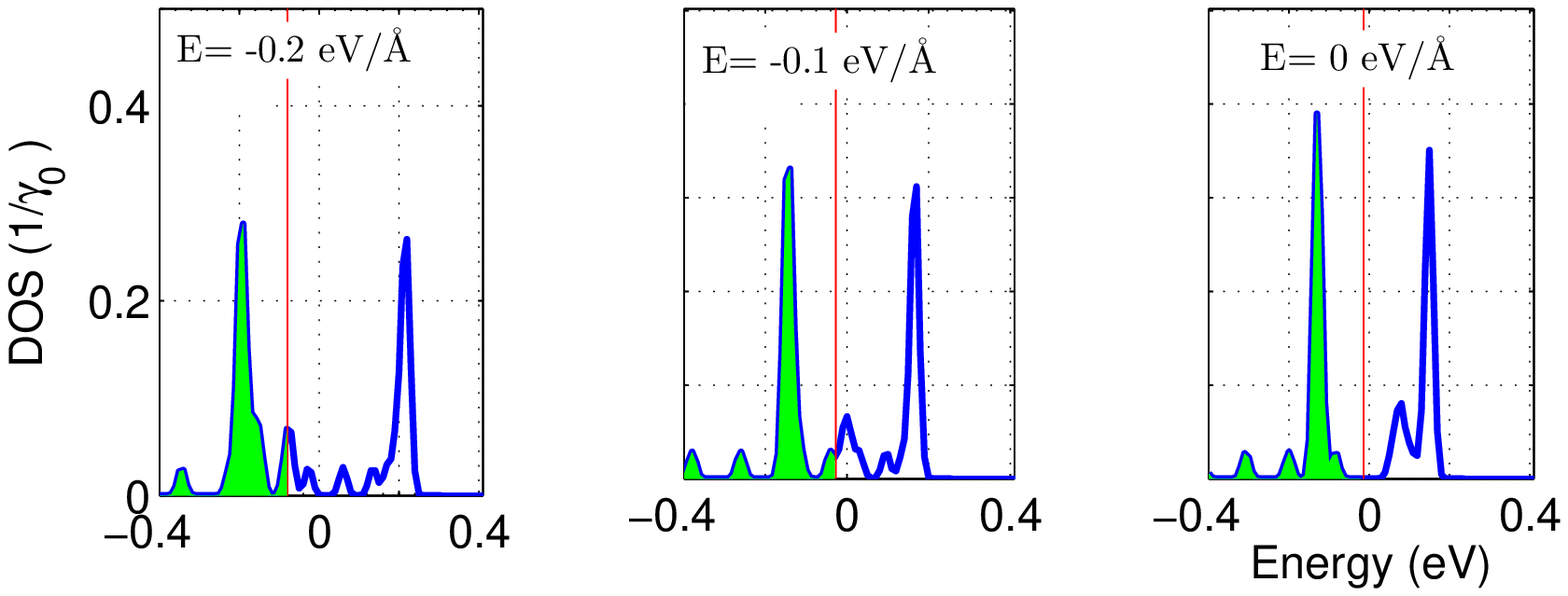}
\end{subfigure}%
\begin{subfigure}{.4\textwidth}
  \centering
\hspace{-.5cm}  \vspace*{.25cm} \includegraphics[width=.8\textwidth]{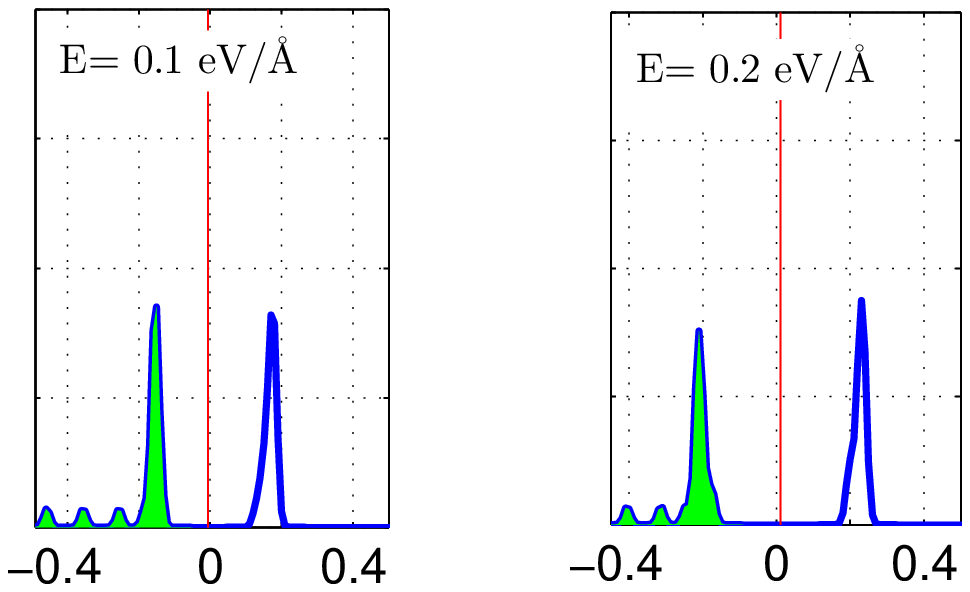}
\end{subfigure}

\begin{subfigure}{.6\textwidth}
  \centering
  \centerline{\includegraphics[width=\textwidth]{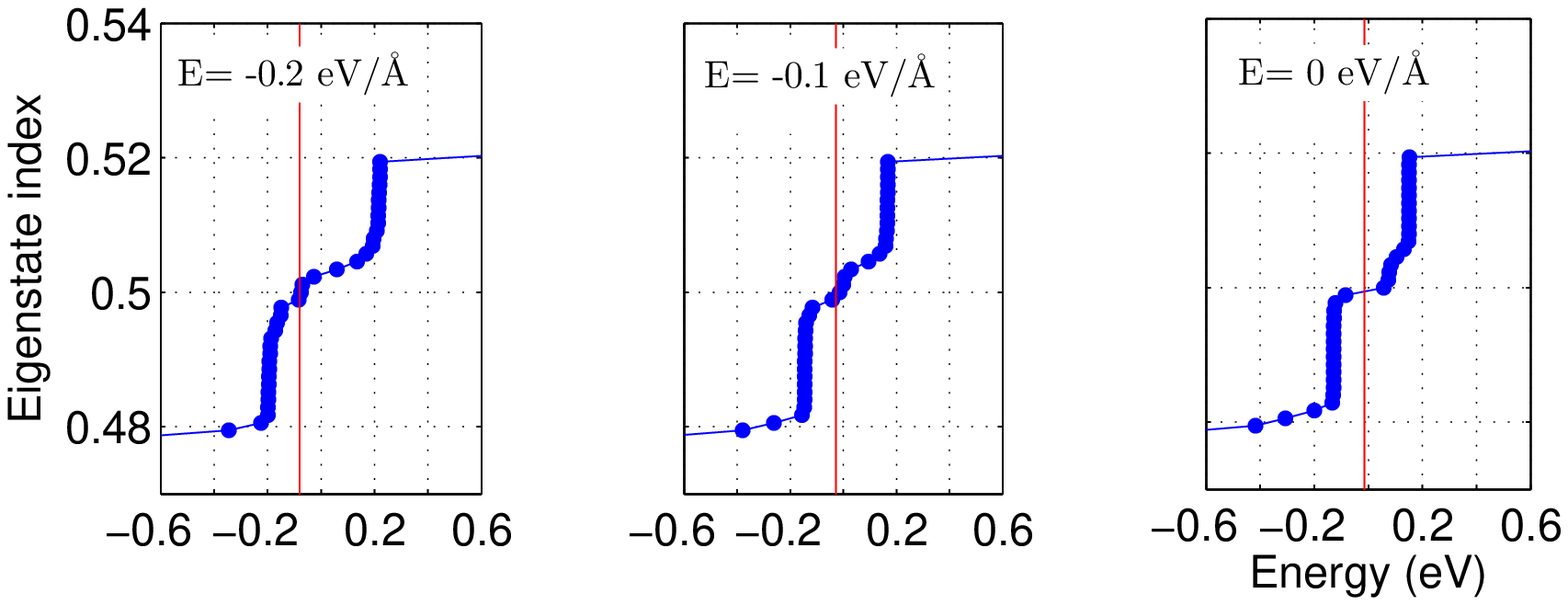}}
\end{subfigure}%
\begin{subfigure}{.4\textwidth}
  \centering
 \hspace{-.5cm}  \vspace*{.19cm} \centerline { \includegraphics[height=3.1cm,width=.88\textwidth]{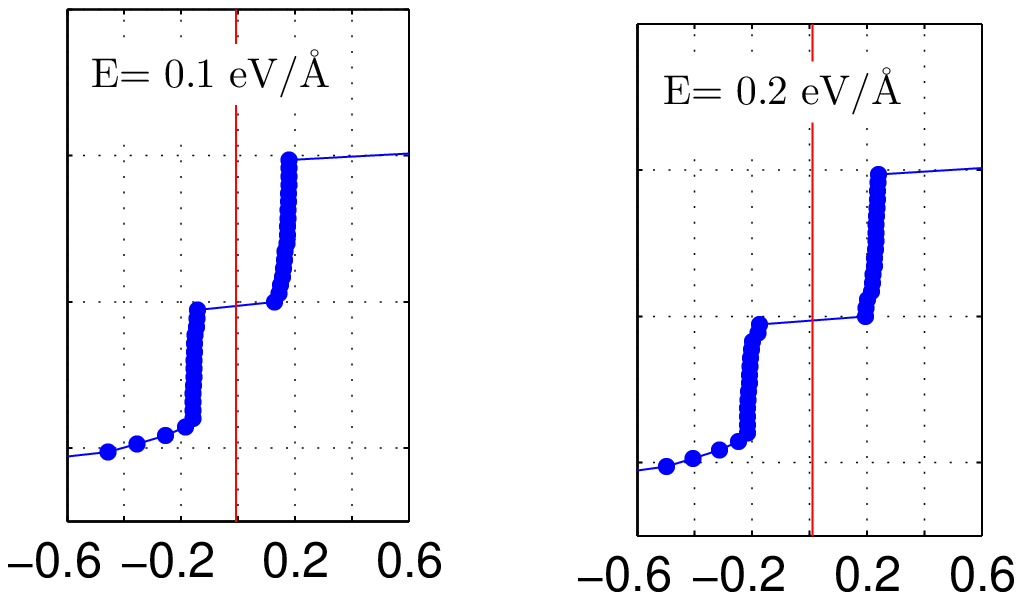}}
\end{subfigure}

\caption{Evolution of DOS (upper panel) and of energy levles spectrum (lower panel) for triangular QDs of bilayer graphene  as function of  applied electric field  in the near-zero energy region.}
\label{Fig4}
\end{figure}

\begin{figure}[htp]
\centering

\includegraphics[width=\textwidth]{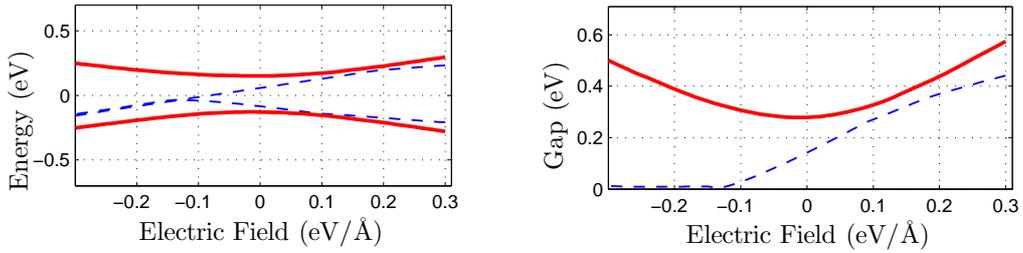}
\caption{(left) Field dependence of peaks of splitted ZES in bilayer graphene (red solid line) and of LOEL and HUEL energies   (dashed blue lines). (right) Corresponding field dependence of peak-to-peak and LOEL-HUEL gaps.}
\label{Fig5}
\end{figure}

Evolution of ZES in triangular QD of bilayer graphene under the action of
electric field was considered in \cite{Guclu} where the  splitting of the highly
degenerate central peak on two gap-separated peaks was predicted. However,
due to used in \cite{Guclu} approximation $\gamma _{1}\neq 0$, $\gamma _{3,4}=0$,
several relevant features were not observed. In more realistic full-$\gamma $
model, ZES are split by parameter $\gamma _{4}$ already at $E=0$ (Fig. \ref{Fig2},
right inset) \cite{Hazem} and, as demonstrated by Fig. \ref{Fig4}, the two-smeared-peak pattern of
DOS just evolves with further extension of the peak-to-peak separation
(Fig. \ref{Fig5}). This peak-to-peak gap should be distinguished from the gap between
the highest unoccupied electronic level (HUEL) and the lowest occupied
electronic level (LOEL). The field evolution of
the latter (Fig. \ref{Fig5}) is not symmetric with respect to the field direction since not
all the edge atoms in the upper plane have their $\gamma _{4}$-partners in
the lower plane and the system is not perfectly symmetrical with respect to
the  mirror reflection.\

\section{Discussion}

Described above possibility of  field-manipulation by electronic properties of Silicene and bilayer graphene QDs
 is provided by the field-dependent
distribution of the edge-localized ZES inside the size-quantized gap and
therefore is quite different from the suggested previously mechanism of the
band-gap tuning in the bulk state \cite{Tsai2013,McCann,Castro2007,CLiu2011}. Moreover, as can be seen from the
previous section, the evolution of ZES is very sensitive to sample geometry.
This certainly creates the wealth of possibilities for design of QDs with
required tunable electronic parameters but poses the query concerning their
stability with respect to the cluster shape change. The fortunate answer is
coming from analysis of the factors affecting ZES behavior. Whereas the
value size-quantized gap with ZES in the middle depends on QD dimmensions,  the
HUEL-LOEL separation, provided by sporadic distribution of individual
levels, is just the function of structure of the edge termination. More
importantly however that such remarkable effects as (i) field-provided displacement
of central peak in triangular Silicene QD, (ii) field-induced gap in
quasi-continuum edge-localized-states spectrum in hexagonal Silicene QD and
(iii) field evolution of peak-to-peak gap splitting of ZES in bilayer graphene QD
are caused by the inter-plane coupling parameters $\gamma _{i}$ and
therefore  are less sensitive to the geometry variation. Our numerical
calculations indeed demonstrate the perfect scalability of these properties, making them a universal feature valid even for ensemble of
similar-shape clusters.

Tight-binding calculations presented here do not take into account the
cooperative Coulomb electron correlations  \cite{Guclu} and another gap-generating effects, 
like e.g. doping \cite{Barrios},  vacuum fluctuations \cite{Kibis} etc. Full account of
these effects under the action of tuning field is the challenging problem.
Another interesting problem is the evolution of the energy spectrum under
the influence of both electric and magnetic field, especially if the last
one exceeds the quantum limit. Generalization of obtaining results for the
multilayer carbon clusters can possibly be useful to explain the
unconventional  tiny magnetic properties of graphite in the ultra-quantum
regime \cite{Kopelevich}.  

This work was supported by the Egyptian mission sector
and by the European mobility FP7 Marie Curie program ITN-NOTEDEV.

\end{document}